# Ontology Assisted Query Reformulation Using the Semantic and Assertion Capabilities of OWL-DL Ontologies

Kamran Munir, Mohammed Odeh and Richard McClatchey

*CCS Research Centre, CEMS Faculty, University of the West of England, ColdharbourLane, Frenchay, Bristol BS16 1QY, UK*

*Email: Kamran.Munir@cern.ch, {Mohammed.Odeh and Richard.McClatchey}@uwe.ac.uk*

**Abstract**

*End users of recent biomedical information systems are often unaware of the storage structure and access mechanisms of the underlying data sources and can require simplified mechanisms for writing domain specific complex queries. This research aims to assist users and their applications in formulating queries without requiring complete knowledge of the information structure of underlying data sources. To achieve this, query reformulation techniques and algorithms have been developed that can interpret ontology-based search criteria and associated domain knowledge in order to reformulate a relational query. These query reformulation algorithms exploit the semantic relationships and assertion capabilities of OWL-DL based domain ontologies for query reformulation. In this paper, this approach is applied to the integrated database schema of the EU funded Health-e-Child (HeC) project with the aim of providing ontology assisted query reformulation techniques to simplify the global access that is needed to millions of medical records across the UK and Europe.*

## 1. Introduction

### 1.1 The problem in general and motivation

Information technology today has been widely adopted in modern medical practice, especially in the support of data management. However little has been achieved in the use of computational techniques to exploit integrated medical information in research. In recent years, there has been a substantial increase in the volume and complexity of data and knowledge available to the medical research community. To enable the use of this knowledge in clinical studies, users generally require an integrated view of medical data across a number of data sources [1]. Clinicians, who are mostly the end users of medical data analysis systems, are normally unaware of the storage structure and access mechanisms of the underlying data sources. Consequently, they require simplified mechanisms for generating queries.

The Health-e-Child (HeC) project [2] aims to develop an integrated platform for European paediatrics, enabling data integration between spatially distributed clinicians and bringing together information produced in different departments or multiple hospitals. The emphasis of the HeC data integration process is on providing "universality of information." Its cornerstone is the integration of information across biomedical abstractions, whereby all layers of biomedical information can be 'vertically integrated' [3] (i.e. integration across cellular, organ, disease, patient and population data). The approach advocated in this paper surrounds the provision of access to an HeC Integrated Data Model [4] plus semantics-driven and transparent query services using manually developed description logic based ontologies. In this regard a framework has been previously presented in [5] which provides transparent query services to access the data.

### 1.2 Health-e-Child query reformulation services

The work presented in this paper exploits the semantic relationships and assertion capabilities of an OWL-DL based ontology in order to capture the domain knowledge and to provide query formulation and reformulation services to the clinicians and their medical applications. To this end a query reformulation system has been developed as middleware between the client applications and distributed data sources (as shown in figure 1). This query reformulation system parses the query and interprets the meaning of the end-user's query terms. In the case where the client request is not automatically resolved or the end-user does not really know what he/she is looking for (or how to ask for available information), the system allows him/her to see and interpret such information. Both of these features enable the construction of a meaningful query.

The process of parsing and interpreting the meaning of the query terms is enabled by the use of source metadata information and domain knowledge that is defined in terms of ontological concepts. These concepts are classified within the internal structure of the ontology. The ontological information is then used for situation-based information querying. Once the new query has been formulated the query is then executed transparently. Users are also able to search and modify previously generated queries.

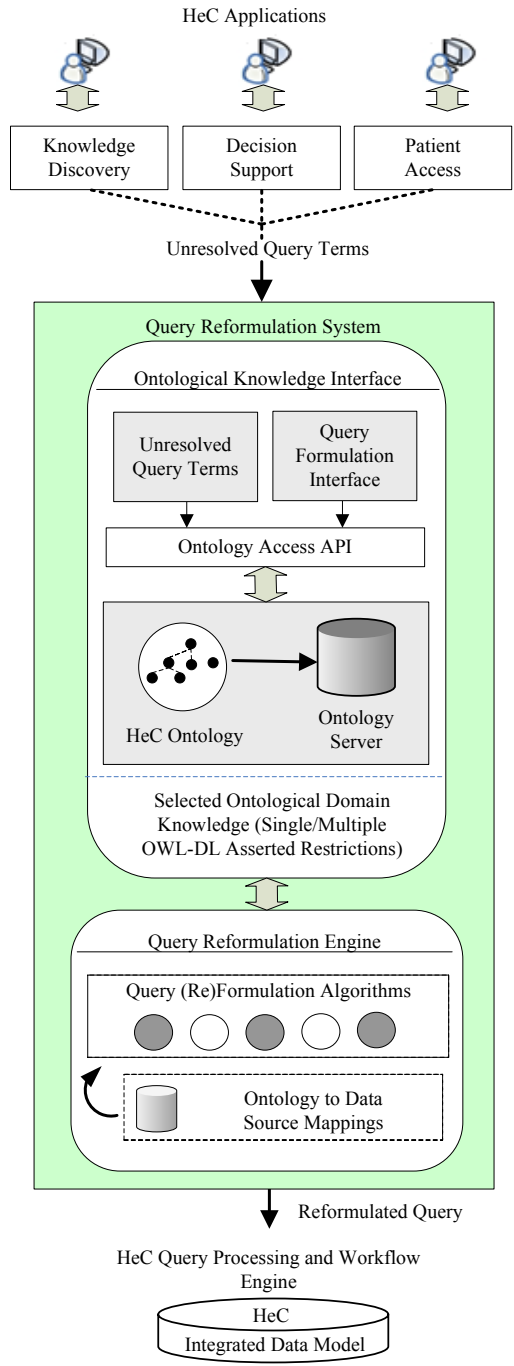

Figure 1: Query Reformulation System Architecture in Health-e-Child.

## 1.1. Query Reformulation and Ontologies

An ontology generally represents a shared, agreed and detailed model (or set of concepts) of a problem domain. One advantage in the use of ontologies in the HeC system is their ability to resolve any semantic heterogeneities that are present within the data. Ontologies can define links between different types of semantic knowledge. Hence, they can aid particularly the resolution of terms for queries and other general search strategies, thus improving the search results that are presented to clinicians. The fact that ontologies are machine-processable and human understandable is especially useful in this regard [6].

There are many biomedical ontologies in existence although few, if any, that support query reformulation over 'vertically integrated' data. The example below explains how ontologies can be used in formulating a query. Due to space limitations, it is not practical to describe the complete HeC database scheme. Thus, as a running example in this paper, we use the following small subset of the database relations from our Patients' database, the complete HeC integrated data model can be found in [4].

(1) patients_data (patient_id, clinical_test_name, clinical_test_value, description, ...)
(2) clinical_test (clinical_test_id, clinical_test_name, ....)
(3) clinical_test_values (id, clinical_test_id, clinical_test_value, ...)
(4) patient (id, description, ...)

The rows in the table *clinical_test* store all the possible clinical tests that can be taken for a particular patient. The *clinical_test_values* table stores all the possible clinical test results for any particular clinical test. The *patients_data* table references the patients, their clinical tests and results of medical tests.

As mentioned previously, ontologies can aid in the area of query reformulation. One example is when a query is reformulated according to the HeC ontology with the concept of 'Astrocytoma Tumor' (a form of tumor in the brain). The user may enter a query into the system stating "Give me all MRI scan images of brains for children with an Astrocytoma Tumor disease in a specific age group". This query cannot be fully resolved by the HeC data model because there is no direct information available in the databases that matches with the term 'Astrocytoma Tumor'. Here the query reformulation system receives a simple input into the system as 'Astrocytoma Tumor', the system then extracts all of the clinical tests and related values that confirms the possibility of Astrocytoma Tumor disease in the brain.

The system uses the power of the HeC ontology to determine that clinical test results for example *headaches, double_vision, orthopedic_sequelea* with values *true, true, severe_symptomatic* are the related clinical tests for Astrocytoma Tumor. Hence, the system not only returns the result as a set of related clinical tests but also returns the respective reformulated query to access this information according to the underlying data model. Such requirements which require resolution of terms from the query reformulation system can occur frequently. Examples of queries with similar requirements include: (1) when a clinician wants to study a particular disease (2) when a clinician wants to study patients who are suspected of having a particular disease and (3) when a clinician wants to compare a patient's medical data with patients who are having a particular disease but also have some other disease e.g. bacterial, congenital or viral infections etc.

The subset of the above mentioned HeC patients' database does not contain information that is directly related to some of the above mentioned terms in queries. Therefore, in order to retrieve the desired query results in the absence of a query reformulation system, clinicians are normally required to perform all of the following operations:

1. To describe the clinical tests that are related to the study of particular diseases;
2. To describe all clinical tests and associated results that need to be "*TRUE*" for one particular patient to be selected as a suspect for particular diseases;
3. To understand how clinical test data is stored in the integrated HeC database; and
4. To write complex SQL queries to select the desired information.

A user may want to write a much more complex query by involving any number of comparisons using *union, intersection, equivalence or negation* operations. Current developments in the field of ontology languages allow ontologies to be more expressive when ontological information is used to formulate complex queries. To achieve this, generic query reformulation techniques have been developed that interpret ontological knowledge to reformulate queries or to assist the users in formulating their queries.

The remainder of this paper is organised as follows. Section 2 presents related work in this area. Section 3 introduces the Query Reformulation system architecture. This section further focuses on the ontological representation for reformulating queries, and discusses the ontology-relational translations that provide (relational) query reformulation services. Finally, Section 4 concludes the paper and considers the current status of research and directions for future work.

## 2. Related Work

Currently, there are several tools available that can transform relational databases into ontologies. DataGenie [7] is a plug-in for Protégé [8] that imports data from a relational database to an ontology. Similarly, related work has been carried out by [9], [10] and [11] on the transformations between relational databases and ontologies. These mappings are fairly trivial: each table maps to a class, each column to a data type property and each row to an instance. In addition, the foreign key columns are used to link an instance of a class to instances of another class. In this research, while using domain ontologies to reformulate relational queries, some of the basic rules to generate domain ontology from relational schema remain the same. However, our relational schema to ontology transformation is different in the sense that 'all' relational data are not transformed and then stored as ontology instances. In addition, ontologies are used to assist in the provision of database querying services to the end users and their applications.

The work presented in [12] supports the specialisation or generalisation of the base or filler concepts to build database specific queries interactively. However, that querying facility cannot generally be applied to queries where the corresponding data is not directly available as ontology instances in the respective data sources. Therefore, implementing such a querying facility in the situation where there are legacy data sources would require significant changes to the database schema. For example, such legacy data sources could involve the medical records of patients (as in HeC*)*. In these cases it is likely to be difficult to manipulate the database schema because of the huge database volume, the associated security protocols and the legacy applications that access them. However, in our approach other than the metadata information, no data is stored as ontology instances or directly linked to ontological concepts. Therefore, no manipulation of the data at the schema level is required. The database metadata is defined within the domain ontology without limiting user queries to the specialisation or generalisation of ontological concepts.

Some of the existing ontology-based information retrieval approaches use RDF [12], [13], [14] and [15] structures which, although yielding schema information, provide insufficient knowledge for query reformulation. These approaches also lack the details of what needs to be included in the ontology from the data sources along with the domain knowledge to drive the process of query reformulation. The focus of these approaches (for example [15]) remains towards interactive query generation through nondirected graphs supporting multiple natural languages. Furthermore, considerable work has been carried out by [16] in addressing the problem of data integration and the interoperation of heterogeneous XML sources using an ontology-based framework, where a global ontology is generated and expressed in an RDF Schema (RDFS) [17]. This system depends on an ontology to define the set of terms that can be used in a query. To query data, users need to be familiar with the overall terms and relationships in the ontology.

This can sometimes be problematic for users who do not fully understand the system and thus face difficulty while navigating through large ontologies to select the appropriate terms. Moreover, in this approach, all data sources need to provide the same view of the information, otherwise finding the minimal ontology commitment becomes a difficult task.

Unlike the approaches in [18], [19], [20], [21] and [22] our system does not store all data from a data-source as part of the ontology or link it directly with ontology concepts. Often it is not practically feasible to store all data as part of a certain domain ontology especially for systems with large amounts of data. The data that is stored as part of the ontology needs to be loaded in memory to perform *Select* query operations. Furthermore, this may become both a complex and time consuming activity in directly linking all data with associated ontological concepts. Most of these approaches have used RDF as an ontology development language. However RDF is too weak to describe resources in sufficient detail since it lacks localised range and domain constraints and there is no support for cardinality constraints. In this research OWL-DL is the ontology development language that is used to specify the concepts with related assertions that drive the process of query reformulation, since it has greater support for expressing semantics when compared to RDF and RDFS. Mappings are defined only between the data source schemas and the basic ontology structure.

Work has been carried out in [23] to map a domain ontology to a domain conceptual data model. In this research several mapping rules have been proposed that guide the transformation from domain ontology to conceptual schema. One of these mapping rules describes the transformation of ontology properties to entities-attributes in the conceptual model. In this paper, this rule has been extended to define mappings between an OWL ontology to a data source schema. In the multiple ontology approach, each data source is described by its own (local) ontology separately. Instead of using a common ontology, local ontologies are mapped to each other. For this purpose, additional representation formalisms are necessary for defining the inter-ontology mappings. The system presented in [13] is one example of such an approach.

A database relational schema provides a logical map of the information content of the database along with related semantic data control assertions, following the relational model. On the other hand, although ontology schemas share the conceptualisation aspects of relational schemas, the ontology model specifications and, in particular OWL ontologies, (used in this research) are based on Description Logic theory [24] and are referred to as OWL-DL. In order to represent a relational data model in OWL-DL, respective transformations of the relational model to DL remain a critical requirement in order to achieve consistency and completeness of these transformations. In relation to this, work has been carried out in [25] which describes the relationship between entities in the ER model and DL theory. In this research some of the basic ontology to DL transformation rules are employed and extended to handle the requirements for reformulating database queries.

## 3. The Query Reformulation System

The query reformulation system reported in this paper (the shaded box in Figure 1) has two major subsystems: (1) the Ontology Knowledge Interface and (2) the Query Reformulation Engine. The Ontology Knowledge Interface subsystem is composed of three components: (a) an ontology creation process to assist in reformulating queries, (b) an ontology server, and (c) an ontology assisted query reformulation process. The Query Reformulation Engine is composed of (a) query reformulation algorithms and (b) ontology to data source mappings.

### 3.1 The Ontological Knowledge Interface

As a first step towards ontology assisted query reformulation, an OWL-DL ontology is created which stores database metadata information within the basic ontology structure. In order to support the re-use, maintainability and evolution of the ontology, a traditional iterative process [26] is adapted for ontology engineering consisting of ontology modeling and ontology validation. In this regard, the metadata from the HeC integrated data model is mapped into disjoint independent trees which are recombined into an ontology using definitions and axioms to represent the relationships in an explicit fashion.

The main elements of a relational database include relations (tables), columns, and constraints (assertions). Equivalently, the ontological model includes classes, properties, assertions and other semantics. However, for the purpose of query reformulation our approach does not require the domain ontology to include all constructs of the relational model. The domain knowledge is expressed in terms of ontology property assertions that need to be consistent with the basic ontology structure. It is also possible to include the domain knowledge from widely available domain ontologies. The mapping rules were developed and have been presented in [5]; that paper explains what needs to be included in the ontology for the purpose of query reformulation.

In our system either the client applications or the user (through an interactive GUI) interacts with the ontology knowledge interface layer to describe the query terms that cannot be automatically resolved from the data sources. The ontology knowledge interface provides access to ontological concepts classified within the internal structure of the ontology.

### 3.1.1 Ontological Representation

This section explains how a subset of the patient's database metadata, used to drive the process of generating queries, is represented in an OWL-DL ontology. Both the domain knowledge and the metadata of the HeC data model are stored in the ontology; a small subset of this metadata is shown here:

*DB relation: clinical_tests*

| Id | name |
|----|------|
| 1  | HEADACHES |
| 2  | DOUBLE_VISION |
| 3  | THROMBOSIS_SEQUELEA |
| 4  | ORTHOPEDIC_SEQUELEA |
| 5  | BACTERIAL_INFECTION |
| …  | … … … |

*PK: id*

*DB relation: clinical_test_values*

| Id | ct_id | ct_value |
|----|-------|----------|
| 3  | 2     | TRUE |
| 4  | 2     | FALSE |
| 5  | 3     | ABSENT |
| 6  | 3     | MODERATE_SYMPTOMATIC |
| 7  | 3     | ASYMPTOMATIC |
| 8  | 4     | SEVERE SYMPTOMATIC |
| 9  | 4     | LIFE THREATENING |
| …  | …     | … … … |

*PK: id, ct_id*
*FK: ct_id reference clinical_tests(id)*

The *clinical_test_values* table stores the possible clinical test results for any particular clinical test id *(ct_id)*. Here *ct_id* (clinical test id) is referenced using the *clinical_test* table. Firstly, a *clinical_tests* class is added to the ontology that contains all of the clinical tests. These would include for example, *headache, double_vision, thrombosis_sequelea, orthopaedic_sequelea* as subclasses, and containing only one instance for each class. The second class, namely *clinical_test_values,* has been defined as a (common) parent class to hold all possible clinical test values for each clinical.

Due to the fact that patients' clinical tests can hold any type of result set values for each clinical test (e.g. *Boolean, String, number etc),* the further subclasses of *clinical_test_values* (e.g. *headaches_values, double_vision_values, thrombosis_sequelea_values and orthopedic_sequelea_values)* have been created. Each of these subclass concepts contains individuals; some of them are shown in Figure 2. We define *clinical_tests* and *clinical_test_values* classes as disjoint so that an individual (or object) cannot be an instance of more than one of these two classes.

Secondly, the object properties, namely *hasClinicalTestName,* and the sub-properties of object property *hasClinicalTestValue* are added. In order to provide a two-way search capability through query reformulation algorithms, these sub-properties have a corresponding inverse property. If a property links individual 'a' to individual 'b', then its inverse property links individual 'b' to individual 'a'. For example, the *Clinical_Tests* individual '*orthopaedic-_sequelea*' is linked with the individuals: *asymptomatic, severe_symptomatic* and *moderate-_symptomatic* with the property *hasOrthopaedic-SequeleaValue*. But, OWL's inverse property *isValueOf* links the individuals *asymptomatic, severe_symptomatic* and *moderate_symptomatic* with the concept *orthopaedic_sequelea*.

If the end users (i.e. clinicians) are accessing the query reformulation system for interactive query generation then during the whole process the users are guided to select the next applicable ontology concept with the corresponding individuals or values. To achieve this task each of the ontology properties has a domain and a range specified. Object properties link individuals from the domain to individuals from the range. For example, the sub-properties of *hasClinicalTestValue* link individuals belonging to the class *Clinical_Tests* to individuals belonging to the *Clinical_Test_Value* class. This is applied to all of the properties available in the ontology, for example the domain of the *hasOrthopaedic-SequeleaValue* property is *Orthopaedic_Sequelea* and the range is *Orthopaedic_Sequelea_Value* as shown (as property links) in Figure 2.

Once the properties with domains and ranges have been defined, then specific class instances are associated with other instances using these object properties. These property links are used to capture the user search criteria within the ontology concepts, or independently from the domain ontology if it is to be further utilised by other users.

Once the basic structural elements of the domain ontology have been defined they are further enriched with domain knowledge. The domain knowledge is expressed in terms of OWL-DL property assertions that need to be consistent with the basic ontology structure. We store this domain knowledge as ontology concepts. In this way the consistency of the domain knowledge with ontology concepts is verified using an Ontology Reasoner [27]. Concept restrictions are used to describe conditions for the selection of records that match some given criteria. These restrictions could be either singular or complex involving many conditions. For example, all conditions must match for a patient record to be selected as a member of a particular disease type (as shown in Figure 3). The query reformulation engine uses these restrictions to reformulate queries by translating the DL constructs into relational queries.

Restrictions in OWL fall into three main categories: *Quantifier, Cardinality* and *hasValue*. *Quantifier* restrictions are used when a restriction is to be placed on an individual to make it a member of a particular class. The *Cardinality* restrictions are used to describe the cardinality of relationships of an individual with the other individuals or datatype values. For the purpose of reformulating queries, the OWL-DL property hasValue restrictions are utilised. The ontology concepts describing a particular *disease study* (as in Figures 2 and 3) embody the associated domain knowledge as well as the search criteria.

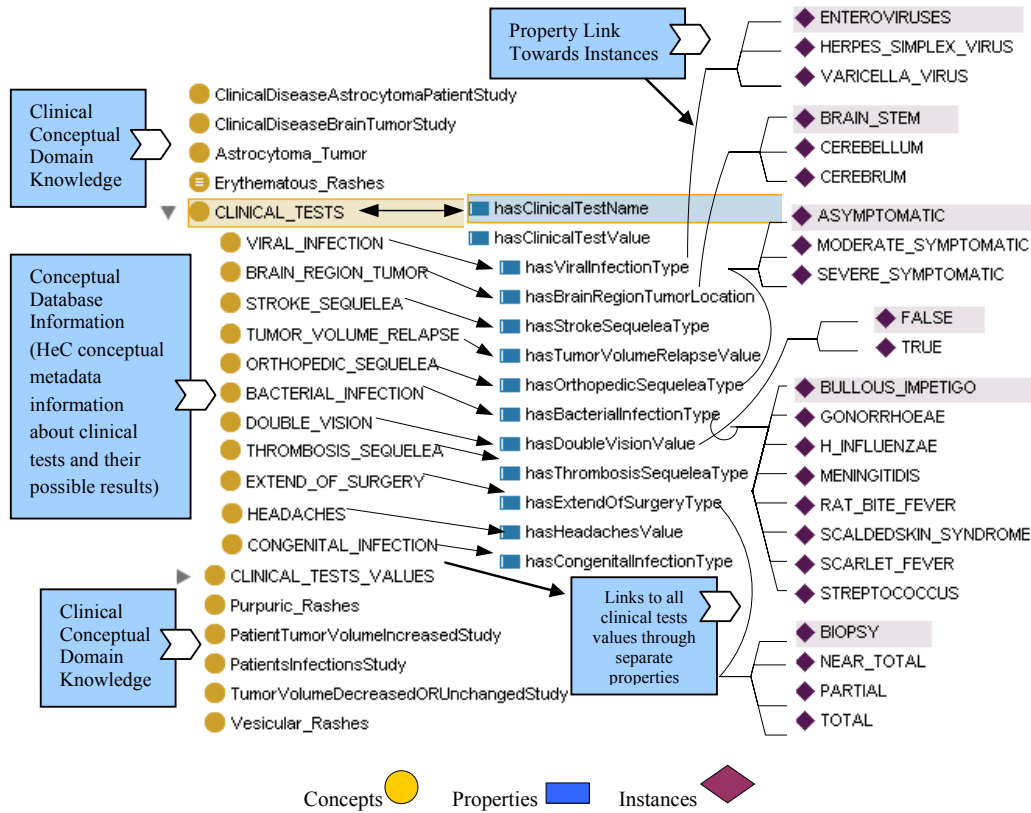

Figure 2: An example of ontological knowledge representation

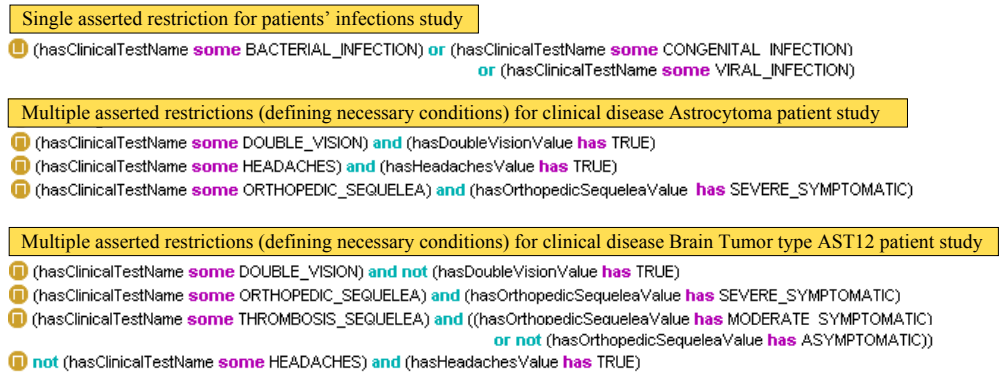

Figure 3: An example of ontology property assertions that drive the selection of relevant records after processing by the query reformulation engine

### 3.1.2 The Ontology Server

Once the ontology has been defined it is then processed and stored in a database. The ontological knowledge interface (as shown in Figure 1) interacts with the ontology server to retrieve the ontological information. This information is then used to assist the users in defining their search criteria (if required) and to generate reformulated database queries by receiving unresolved query terms from client applications. The consistency of the domain knowledge with ontology concepts can be verified using an ontology Reasoner (e.g. FaCT++, Racer) [27]; however, if the domain knowledge is to be accessed from a database then this requires the implementation of a consistency check mechanism to ensure coherence with the ontology. This domain knowledge is used by the query reformulation algorithms to reformulate queries conforming to the schema(s) of the underlying data sources.

### 3.1.3 The Ontology Assisted Query Formulation Process

As depicted in Figure 1 the ontology knowledge interface can receive requests from the client applications as well as from the end users. In situations when end users are directly accessing the ontology knowledge interface, the user could define a new search criterion or select from the existing domain knowledge to formulate a query. The users' search criteria are described using the ontology property restrictions, concepts/sub-concepts and instances. An individual must satisfy all the conditions that are specified as ontology property restrictions to be a member of any named concept. These restrictions could be either singular or complex ones involving many conditions. For example, restrictions are used to define conditions for the selection of relevant patient records that match a given criteria. A GUI interface, the so-called *"Ontology Assisted Query Formulation Wizard"*, is provided for this task which guides the user in defining the search conditions by making full use of the supporting domain ontology.

For example, in order to generate the query (query-1) where a user wants to retrieve clinical test data for each patient related to the study of Brain Tumor Disease-X, the selection criterion is described as OWL property assertions, e.g. by using an ontology property namely *"hasClinicalTestName"*. Once defined, the whole search criteria are saved as a new ontological concept for example, *brain_tumor_disease_x_study*. In this example, it is assumed that the *double_vision, headaches* and *orthopaedic_sequelea* are the clinical test names related to brain tumor disease *x*. This information is described and saved as follows:

Concept Name: Brian_tumor_disease_x_study
(OWL expression)
*hasClinicalTestName some* DOUBLE_VISION <u>union</u>
*hasClinicalTestName some* HEADACHES <u>union</u>
*hasClinicalTestName some* ORTHOPAEDIC_SEQUELEA

For the situations where the ontology knowledge interface receives requests from the client applications to reformulate queries for the 'unresolved query terms', the ontology access API accesses the ontology to extract relevant OWL-DL assertions. For example (query-2), when a user wants to retrieve information about patients who are suspected to have a particular Brain Tumor Disease-X (e.g. 'Astrocytoma Tumor') the system receives the query term 'Astrocytoma Tumor' and extracts all of the clinical tests and related values that confirms the possibility of the Astrocytoma Tumor disease in the brain. The system uses the HeC ontology to determine that the clinical test results for example *headaches, double_vision, orthopaedic_sequelea* with values *true, true, severe_symptomatic* are the related clinical tests for Astrocytoma Tumor. All these conditions need to be satisfied in order to indicate a suspected case of brain tumor disease *x*:

Concept Name: <u>Brian_Tumor_Disease-X_Suspects</u>
(OWL expression)
*{AllValuesRestriction*
 *(hasClinicalTestName some* DOUBLE_VISION
*Intersection*
 *hasClinicalTest BooleanValue has* TRUE)
*AllValuesRestriction*
 *(hasClinicalTestName some* HEADACHES
*Intersection*
 *hasClinicalTestBooleanValue has* TRUE)
*AllValuesRestriction*
 *(hasClinicalTestName some* ORTHOPEDIC_SEQUELEA
*Intersection*
*hasClinicalTestStringValue has* severe_symptomatic)}

These query conditions could, of course, be more complex since it could involve multiple ontology assertions using a mixture of *union, intersection, equivalence and negation* operations even within each property restriction. The query reformulation for such cases involves the handling of all different situations. In the next section, we show how these object property assertions, domain knowledge and associated database mappings are utilised to reformulate the respective query.

### 3.2 The Query Reformulation Engine

The Query Reformulation Engine is composed of query reformulation algorithms and ontology to data source mappings. The query reformulation interface passes the extracted relevant ontological information to the query reformulation engine. The query reformulation algorithms interpret and transform the OWL Description Logic constructs into corresponding relational (SQL) queries. The *mappings* table is created automatically during the ontology processing that stores the information about ontology property links, database name, table name, column name, primary and foreign keys. Once created this mapping table only contains the information about ontology properties, which is then updated with the database metadata information.

For example, from the selection conditions defined for query 1 (as discussed in section 3.1.3), it can be deduced that the *double_vision, headaches* and *orthopaedic_sequelea* are the clinical tests related to *Brain Tumor Disease-X Study* and defined with the *'OR'* condition. Here the *'OR'* condition for all parts of the test condition implies that there is a *'UNION'* operation within each test condition for the data in *Patients'* database. Finally, the formulated query in this case will retrieve the patient data from the *Patient_Information* database view/table where *patient_information. clinical_test_name* matches any of the following values {*double_vision, headaches, orthopaedic_sequelea*}. Query 2 is more complex than query 1 and from the previously defined selection conditions for query 2, we deduce that the following asserted restrictions are indicative of Brain Tumor Disease-*X*:

| Ontology Restrictions | Ontology Properties | Test Conditions |
|---|---|---|
| **Condition-1** *(condition 1A)* | *hasClinical TestName* | DOUBLE_VISION |
| **AND** *(condition 1B)* | *hasDouble VisionValue* | TRUE |
| **Condition-2** *(condition 2A)* | *hasClinical TestName* | HEADACHES |
| **AND** *(condition 2B)* | *hasHeadaches Value* | TRUE |
| **Condition-3** *(condition 3A)* | *hasClinical TestName* | ORTHOPAEDIC_ SEQUELEA |
| **AND** *(condition 3B)* | *hasOrthopedic SequeleaValue* | SEVERE_ SYMPTOMATIC |

Table 1: Ontology asserted restrictions for the Brain Tumor Disease-X suspects.

Here the clinical test variables *double_vision, headaches* and *orthopaedic_sequela* with the clinical test values *'true'*, *'true'* and *'severe_symptomatic'* respectively are defined as multiple restrictions for *Brain_Tumor_Disease_X_ Suspects*. Also the multiple asserted restrictions and *'AND'* conditions for each *hasValue* property within each restriction implies that there is an *Intersection* of conditions within rows and columns. In this case the reformulated query for query 2 will retrieve all patients that have all *Clinical Tests* recorded for *Disease-X* with specific values for each *Clinical Test* i.e. for *double_vision = 'true', headaches = 'true'* and *orthopaedic_sequelea = 'severe_symptomatic'*.

As described previously (in section 3.1) the property *hasClinicalTestValue* is a parent property of the *hasDoubleVisionValue, hasHeadachesValue and hasOrthopaedicSequeleaValue* objects. In this approach we require mapping definitions only for the parent properties. The following are the mappings for the ontology properties *hasClinicalTestName* and *hasClinicalTestValue* for the *Patient_Information* database view.

*hasClinicalTestName* → *(belongs to) patient_information. clinical_test_name*
*hasClinicalTestValue* → *(belongs to) patient_information.clinical_test_value*

The query2 example becomes more complicated (see below) as there are multiple test conditions with *'OR/UNION'* operations e.g. for clinical test name 'orthopedic_sequelea'.

Concept Name: Brian_Tumor_Disease-X_Suspects:
{*hasClinicalTestName some* DOUBLE_VISION
*hasClinicalTestBooleanValue has* TRUE
*Intersection*
*hasClinicalTestName some* HEADACHES
*hasClinicalTestBooleanValue has NOT* TRUE
*Intersection*
{*hasClinicalTestName some* ORTHOPAEDIC_SEQUELEA
hasClinical*TestStringValue has* SEVERE_SYMPTOMATIC
*Union*
*hasClinicalTestName some* ORTHOPAEDIC_SEQUELEA
hasClinical*TestStringValue has* (SEVERE_SYMPTOMATIC *Union* LIFE_THREATENING*)*}}

| Ontology Restrictions | Ontology Properties | Test Conditions |
|---|---|---|
| **Condition-1** *(condition 1A)* | *hasClinical TestName* | DOUBLE_ VISION |
| **AND** *(condition 1B)* | *hasClinicalTes tBoolean Value* | TRUE |
| **Condition-2** *(condition 2A)* | *hasClinical TestName* | HEADACHES |
| **AND NOT** *(condition 2B)* | *hasClinicalTes tBooleanValue* | TRUE |
| **Condition-3** *(condition 3A)* | *hasClinical TestName* | ORTHOPAEDIC_ SEQUELEA |
| **AND** *(condition 3B)* | *hasClinical TestString Value* | SEVERE_ SYMPTOMATIC ***OR*** *(condition 3C)* LIFE_ THREATENING |

Table 2: An example of complex ontology asserted restrictions for the Brain Tumor Disease-X suspects.

The above form of user query is more complex since it involves multiple ontology assertions involving a mixture of *union* and *intersection* operations within each property restriction. In the next section, we outline the mappings between an ontology model and a relational database model. These mappings provide us with the ground on which we have based (and implemented) our query reformulation algorithms in the query reformulation engine to handle the possible Description Logic expressions to respective Relational Query translations.

## 3.3 Mappings from an Ontology Model to a Relational Model

A relational data model aims at establishing links between user and domain requirements and describes the logical structure and contents of the data. However, it is often necessary to clarify the meaning of the entities and their properties for a specific domain of interest to aid understanding. An ontology is one way of describing these entities along with their properties in the real world [6]. Recently, semantic web ontology languages have been used to express different types of ontologies and associated languages such as OWL to help in modeling the real world more accurately. These ontologies play a significant role in information system modeling and have the ability to represent the conceptual data models using ontological theories [23]. Moreover (as discussed previously in sections 1.3 and 2.0), similar work has also been reported in [10] and in particular the R2O System [11] that describes the mappings between a relational database schema and an ontology.

Figure 4 presents different mapping situations that arise from ontology-to-relational and relational-to-ontology model mapping scenarios and are covered by our query reformulation algorithms (detailed examples of these mappings are reported in [5]). Here the mappings are expressed as a set of correspondences that relate the vocabulary of a relational model (table/relation, column etc) with an ontology model (concept, property etc) and vice versa.

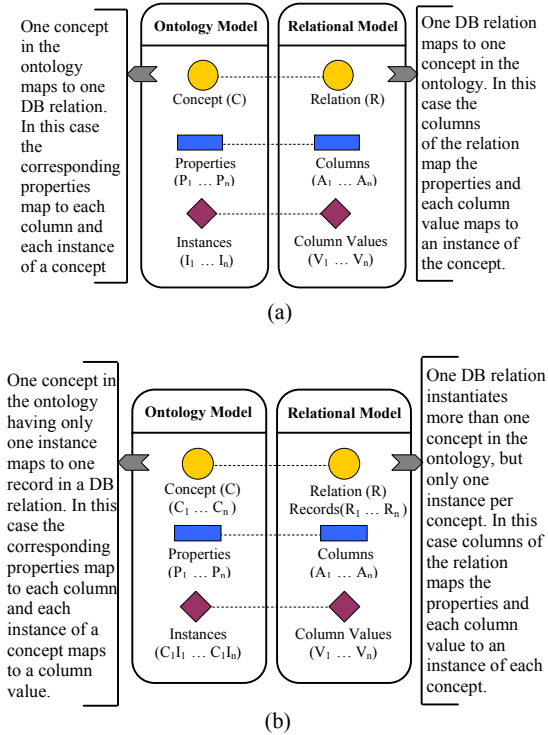

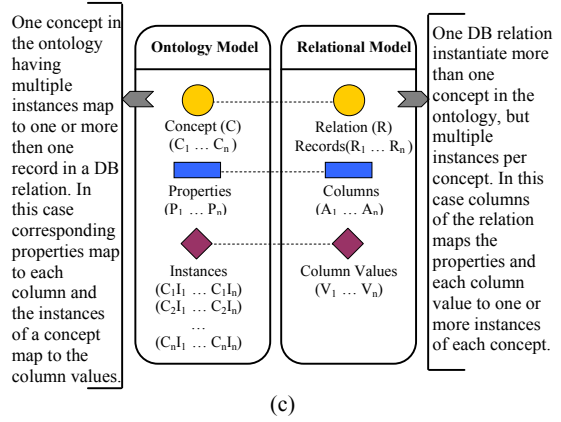

Figure 4(a, b, c): Mappings between an ontology model and a relational model

In the relational database paradigm, a logical data model may be accessed through SQL which is based on the Relational Algebra (RA), whereas OWL-DL is based on Description Logic [24]. Therefore, we base our translations on Description Logic and Relational Algebra, to work with any database that implements the SQL standard.

In DL, a given DL is defined by a set of concepts and a role forming operator. The smallest set propositionally closest to DL is $\mathcal{ALC}$ (Attributive Language with Complements) where the concepts are constructed using *Union, Intersection, allValuesFrom, someValuesFrom and complementOf* written as $\cup, \cap, \forall, \exists, and \neg$, respectively. The 'all' in *allValuesFrom* is the universal qualifier whereas the 'some' in *someValuesFrom* is the extensional qualifier. The *someValuesFrom (hasClass)* and *allValuesFrom (toClass)* constructs are applied on classes or subclasses while specifying classes and restrictions, whereas the *hasValue* is used with instances.

In the next section, we outline the DL to Relational Algebra (RA) translation heuristic rules rather than a formal approach to achieve this translation from OWL-DL ontological queries to relational queries that can be executed by the relational query processor of a relational database management system.

## 3.4 Translation of OWL DL Constructs into Relational Queries

From this point onwards, the following conventions (as per the mappings defined in the previous section) have been used in the translation rules from DL to RA queries.

- ζ represents an ontology or ontology fragment
- $Q_R$ is the formulated query in RA
- R is a database relation/view
- C is the ontology concept/class
- $C_1, C_2 \ldots C_n$ are multiple ontology concepts/classes

- P is an ontology property (mapped to a database column)
- $P_1, P_2 \ldots P_n$ represent multiple ontology properties (mapped to database columns)
- I is an ontology instance (mapped to a database column-value)
- $I_1, I_2 \ldots I_n$ represent ontology concept-instances (mapped to database column values)
- pk_column is the primary key column for a database relation R

### 3.4.1 Translations for the *allValuesFrom* DL construct

The *allValuesFrom* restriction excludes the possibility of further additions for a given property. The "*allValuesFrom*" is interpreted as "*only*", such that saying all values coming from a given class is the same as saying that values may only come from that class. While defining OWL property assertions the "*allValuesFrom*" may be used in the following ways:
(1) Concept (C) having only one instance (I):
   <object-property>*allValuesFrom*(*Class*)
(2) Concept (C) having multiple instances ($I_1 \ldots I_n$):
   <object-property>*allValuesFrom*{*class-instance*
   <space> *class-instance* ......}
For both of these cases the following query reformulation rules are used to generate a relational query.

1. An Ontology assertion with an *allValuesFrom*($\forall$) restriction for a property (P) on a concept (C), implies:
   If 'C $\in \zeta$ | *allValuesFrom* ($\forall$) of P *toClass* C' then the translated relational algebra query will be:

$$Q_R = \sigma_{P_{pk\_column} \neg IN (\pi_{pk\_column} (\sigma_{P<,>C}(R)))}(R)$$

2. If C $\in \zeta$ | *allValuesFrom* ($\forall$) of P *toClass (class-instances)* ($I_1 I_2 \ldots I_n$) then
   (Here {$I_1$ <space> $I_2$ <space>..... $I_n$} are the class instances for property P)

$$Q_R = \sigma_{P_{pk\_column} \neg IN (\pi_{pk\_column} (\sigma_{\neg(P=I_1 \vee P=I_2 \vee ..... \vee P=I_n)}(R))}(R)$$

### 3.4.2 Translations for the *someValuesFrom* DL construct

The "*someValuesFrom*" is interpreted as "*some*", such that the values may come from a given class. The DL to RA translation for a single *someValuesFrom* property restriction on a concept (C), having only one instance (I) is fairly straightforward. For example, in such a case the ontology property maps to the column name and the concept maps to the column value. Therefore, we only present the translations concerned with the more than one occurrence of the *someValuesFrom* property restriction, and with the *Union* or *Intersection* or both operations within each restriction. For such cases the following scenarios (and combinations of all these) can occur:

**Scenario 1**: Multiple *someValuesFrom* ($\exists$) (restrictions), with a *Union* operation within each restriction, and a (similar) property defines each concept.

Such a scenario only occurs for a class having subclasses, and a property defines the parent class as a Range class.
*((<property P> someValuesFrom <class $C_1$>) Union*
 *(<property P> someValuesFrom <class $C_2$>))> .......*
*(<property P> someValuesFrom <class $C_n$>*

Here, the ontology assertions with the *someValuesFrom* ($\exists$) property restrictions for ($C_1$, $C_2 \ldots C_n$) with *Union* operation within each *someValuesFrom* property restriction, imply:
If C $\in \zeta$ | (*someValuesFrom* ($\exists$) of P *some* ($C_1$, $C_2 \ldots C_n$)) then

$$Q_R = \sigma_{P=C_1 \vee P=C_2 \vee .... \vee P=C_n}(R)$$

**Scenario 2**: Multiple *someValuesFrom* ($\exists$), with a *Union* operation within each restriction, and a distinct property defines each concept.
*((<property $P_1$> someValuesFrom <class $C_1$>) Union*
 *(<property $P_2$> someValuesFrom <class $C_2$>))*

If C $\in \zeta$ | (*someValuesFrom* ($\exists$) of P ($P_1 \cup P_2 \cup \ldots \cup P_n$) *some* ($C_1$, $C_2 \ldots C_n$) then
(Here ($P_1, P_2 \ldots P_n$) are the ontology properties for the concepts ($C_1, C_2 \ldots C_n$) having only one instance per class.)

$$Q_R = \sigma_{P_1=C_1 \vee P_2=C_2 \vee ..... \vee P_n=C_n}(R)$$

**Scenario 3**: Multiple *someValuesFrom* ($\exists$), with an *Intersection* operation within each restriction, and a distinct property defines each concept.

Such a scenario only occurs for distinct *someValuesFrom* properties, and a property defines the parent class as a Range class.
*((<property $P_1$> someValuesFrom <class $C_1$>)Intersection*
 *(<property $P_2$> someValuesFrom <class $C_2$>))*

If C $\in \zeta$ | (*someValuesFrom* ($\exists$) of P ($P_1 \cap P_2 \cap \ldots \cap P_n$) *some* ($C_1$, $C_2 \ldots C_n$) then

$$Q_R = \sigma_{P_1=C_1 \wedge P_2=C_2 \wedge ..... \wedge P_n=C_n}(R)$$

**Scenario 4**: Multiple *someValuesFrom* ($\exists$), with the both *Intersection* and *Union* operations.

Such a scenario requires combining both above mentioned (2nd and 3rd) rules.

*(<property $P_i$> someValuesFrom <class $C_i$>)*
*Intersection/Union*
*((<property $P_{1j}$> someValuesFrom <class $C_{1j}$>) Union*
 *(<property $P_{2j}$> someValuesFrom <class $C_{2j}$>))*

If C $\in \zeta$ | ((*someValuesFrom* ($\exists$) of $P_i$ *some* ($C_{1i}, C_{2i}$ … $C_{ni}$) $\cap$|$\cup$((*someValuesFrom* ($\exists$) of $P_j$ ($P_{1j} \cap P_{2j}$ $\cap$… $\cap P_{nj}$) *some* ($C_{1j}, C_{2j}$ … $C_{nj}$)) then

$$Q_R = \sigma_{(P_i=C_{1i} \vee P_i=C_{2i} \vee ..... \vee P_i=C_{ni})\ and\ /\ or\ ((P_{1j}=C_{1j}) \wedge (P_{2j}=C_{2j}) \wedge ..... \wedge (P_{nj}=C_{nj}))}(R)$$

**Scenario 5**: Multiple assertions of *someValuesFrom* ($\exists$).

Such a scenario can occur only with the distinct ontology properties.

<property $P_1$>   *someValuesFrom*  <class $C_1$>
<property $P_2$>   *someValuesFrom*  <class $C_2$> …….
<property $P_n$>   *someValuesFrom*  <class $C_n$>

If C $\in \zeta$ | *multiple assertions* | (*someValuesFrom* ($\exists$) of $P_1$ *some* $C_1$), (*someValuesFrom* ($\exists$) of $P_2$ *some* $C_2$) …... (*someValuesFrom* ($\exists$) of $P_n$ *some* $C_n$) then

$$Q_R = \sigma_{P_1=C_1 \wedge P_2=C_2 \wedge ..... \wedge P_n=C_n}(R)$$

**Scenario 6**: A *someValuesFrom* ($\exists$) restriction with the multiple instances of a concept.

Such a scenario can occur when multiple instances of concept are defined with a *someValuesFrom* property restriction.

<object-property> *someValuesFrom* {class-instance <space> class-instance ……}

If C $\in \zeta$ | *someValuesFrom* ($\exists$) of P *hasClass* (class-instances) $\{I_1, I_2 …. I_n\}$ then
(Here $\{I_1$ <space> $I_2$ <space>….. $I_n\}$ are the class instances for property P)

$$Q_R = \sigma_{P=I_1 \vee P=I_2 \vee ..... \vee P=I_n}(R)$$

### 3.4.3. Translations for the *complementOf* DL construct

The *complementOf* DL construct selects all individuals that do not fall under the specified restriction(s). For a single ontology assertion with only one *complementOf* property restriction the translation is trivial. For example, in such a case the ontology property maps to the column name and the individual maps to the column value with a *NOT equal-to* condition. Therefore, here we only present the translations concerned with the more than one occurrence of the *complementOf (hasValues)* property restriction, and with the *Union* or *Intersection* or both operations within each restriction. For such cases the following three scenarios (and combinations of all these) can occur:

**Scenario 1**: A *complementOf* construct, with a *Union* operation within each *hasValue* property restriction.
*complementOf* (*hasValues of $I_1$ Union hasValues of $I_2$ Union …… Union hasValues of $I_n$*)

If C $\in \zeta$ | *complementOf* (*hasValues* ($\ni$) of P *has* ($I_1$ $\cup$ $I_2$ $\cup$ … $\cup$ $I_n$)) then

$$Q_R = \sigma_{\neg(P=I_1 \vee P=I_2 \vee ..... \vee P=I_n)}(R)$$

**Scenario 2**: Multiple *complementOf* constructs, with an *Intersection* operation within each restriction.
*complementOf* (*hasValues of $I_1$*) Intersection
*complementOf* (*hasValues of $I_2$*) Intersection ……
Intersection *complementOf* (*hasValues of $I_n$*)

If C $\in \zeta$ | <*complementOf*> *hasValues* ($\ni$) of P *has* ($I_1 \cap I_2 \cap … \cap I_n$) then

$$Q_R = \sigma_{\neg(P=I_1) \wedge \neg(P=I_2) \wedge ..... \wedge \neg(P=I_n)}(T)$$

**Scenario 3**: Multiple *complementOf* constructs, with the both *Intersection* and *Union* operations.

Such a scenario requires combining both above mentioned ($1^{st}$ and $2^{nd}$) rules.
(*complementOf* (*hasValues of $I_{1i}$*) Intersection ……
Intersection *complementOf* (*hasValues of $I_{ni}$*))
Intersection/Union
(*complementOf* (*hasValues of $I_{1j}$* Union … Union *hasValues of $I_{nj}$*))

If C $\in \zeta$ | (<*complementOf*> *hasValues* ($\ni$) of $P_1$ *has* ($I_{1i} \cap I_{2i} \cap … \cap I_{ni}$) $\cap$|$\cup$ <*complementOf*> ((*hasValues* ($\ni$) of $P_2$ *has* ($I_{1j} \cup I_{2j} \cup … \cup I_{nj}$)) then

$$Q_R = \sigma_{(\neg(P_1=I_{1i}) \wedge \neg(P_1=I_{2i}) \wedge ..... \wedge \neg(P_1=I_{3i})) \wedge/\vee (\neg(P_2=I_{1j} \wedge P_2=I_{2j} \wedge ..... \wedge P_2=I_{nj}))}(R)$$

### 3.4.4 Translations for the *hasValue* DL construct

A *hasValue(has)* restriction, denoted by the symbol ($\ni$), describes the set of individuals that have at least one relationship along a specified property to a specific individual. Some of the basic translations for *hasValue* property restrictions are almost similar to the scenarios 1, 2 and 3 described previously for the suggested *someValuesFrom* DL construct translation. The only major difference between them is that *hasValue* describes the set of individuals and *someValuesFrom* describes the ontology concept(s). Therefore, in this section we present two further example translations concerned with the more than one occurrence of the *hasValues* property restriction, with the *Union* or *Intersection* or both operations within each property restriction.

**Scenario 1:** Multiple *hasValue* constructs, with the *Union* or *Intersection* or both operation(s).
(((<property $P_1$> *hasValue* <instance $I_{1i}$>) Union
 (<property $P_1$> *hasValue* <instance $I_{2i}$>) Union……)
 Intersection
(<property $P_2$> *hasValue* <instance $I_{1j}$>) ……

If C $\in \zeta$ | (*hasValues* ($\ni$) of $P_{1i}$ *has* ($I_{1i} \cup I_{2i} \cup … \cup I_{ni}$)) $\cap$|$\cup$ (*hasValues* ($\ni$) of ($P_{1j} \cap P_{2j} \cap … \cap P_{nj}$) *has* ($I_{1j}, I_{2j}$ … $I_{ni}$)) then

$$Q_R = \sigma_{(P_{1i} = I_{1i} \vee P_{1i} = I_{2i} \vee ..... \vee P_{1i} = I_{ni}) \text{ and / or}}$$
$$((P_{1j} = I_{1j}) \wedge (P_{2j} = I_{2j}) \wedge ..... \wedge (P_{nj} = I_{nj}))} (R)$$

**Scenario 2:** Multiple assertions of a *hasValue* construct.

Such a scenario can occur only with the distinct properties.

<property $P_1$> hasValue(has) <instance I>
<property $P_2$> hasValue(has) <instance I> ......
<property $P_n$> hasValue(has) <instance I>

If $C \in \zeta$ | *multiple assertions* | (*hasValues* ($\ni$) of $P_1$ *has* I), (*hasValues* ($\ni$) of $P_2$ *has* I) ...... (*hasValues* ($\ni$) of $P_n$ *has* I) then

$$Q_R = \sigma_{P_1 = I \wedge P_2 = I \wedge ..... \wedge P_n = I}(R)$$

### 3.4.5 Translations for the combinations of the *someValuesFrom* (i.e. *hasClass*) and the *hasValue* (i.e. *hasInstance*) DL constructs

As described previously, the "*someValuesFrom*" DL construct is interpreted as "some", such that the values may come from a given class and a *hasValue* restriction describes the set of individuals. In this section, we present the example translations concerning the combinations of both, the *someValuesFrom* (i.e. *hasClass*) and the *hasValue* (i.e. *hasInstance*) DL constructs. For such cases the following three scenarios (and also the all possible combinations of the previously specified scenarios for the *someValuesFrom*, the *hasValue* and the *complementOf* DL constructs can occur:

For the following examples, $P_s$ represents the *someValuesFrom* ($\exists$) and $P_h$ represents the *hasValue* ($\ni$) related ontology properties.

**Scenario 1:** A single restriction with the (combination of) *someValuesFrom* (i.e. *hasClass*) and the *hasValue* (i.e. *hasInstance*) DL constructs.

The *someValuesFrom* and *hasValue* constructs are used together when restrictions are to be placed on the instance(s) (as *hasValue*) of a particular concept (i.e. *hasClass*). The following combination can be interpreted as 'value may come from a class 'C' that matches the instance 'I'. This is particularly useful when a selection condition is to be applied on more then one column of a DB relation/view.

((<property $P_s$ >someValuesFrom <class C>)
Intersection
(<property $P_h$ > hasValue <instance I>))

If $C \in \zeta$ | (*someValuesFrom* ($\exists$) of $P_s$ *some* C) $\cap$ (*hasValue* ($\ni$) of $P_h$ *has* I) then

$$Q_R = \sigma_{P_S = C \wedge P_h = I}(R)$$

In such a case the *Intersection (And)* operation is applied between the *hasClass* and *hasValue* constructs, and both of the conditions need to be true for the selection of a particular record.

**Scenario 2:** Multiple restrictions with the (combination of) *someValuesFrom* and the *hasValue* DL constructs, with a *Union* operation within each combine (*someValuesFrom, hasValue*) restriction.

In such a scenario the conditions are applied to the multiple concepts and their corresponding instances.

(<property $P_s$> someValuesFrom <class $C_1$>
Intersection <property $P_h$> hasValue <instance I>)
Union
(<property $P_s$> someValuesFrom <class $C_2$>
Intersection <property $P_h$> hasValue <instance I>)
Union ................

If $C \in \zeta$ | (($\exists$ of $P_s$ *some* $C_1$ $\cap$ $\ni$ of $P_h$ *has* I) $\cup$
($\exists$ of $P_s$ *some* $C_2$ $\cap$ $\ni$ of $P_h$ *has* I) .....
($\exists$ of $P_s$ *some* $C_n$ $\cap$ $\ni$ of $P_h$ *has* I)) then

(Here '$\exists$' is represents *someValuesFrom* and '$\ni$' represents *hasValues*.)

$$Q_R = \sigma_{(P_s = C_1 \wedge P_h = I) \vee (P_s = C_2 \wedge P_h = I)}$$
$$\vee ..... \vee (P_s = C_n \wedge P_h = I)} (R)$$

**Scenario 3**: *(Multiple Assertions a, b ... n)*

This scenario discusses the translation for the multiple restrictions with the (combination of) *someValuesFrom* and the *hasValue* DL constructs, with an *Intersection* operation within each combine (*someValuesFrom, hasValue*) restriction. This is useful for the cases where it is required to test multiple test conditions that all need to be *'TRUE'*.

a. (<property $P_s$> someValuesFrom <class $C_1$>)
   Intersection
   (<property $P_h$> hasValue <instance $I_{c1}$>)
b. (<property $P_s$> someValuesFrom <class $C_2$>)
   Intersection
   (<property $P_h$> hasValue <instance $I_{c2}$>)
c. .........
.

(For this scenario $I_{c1}$ ... $I_{cn}$ represents the corresponding instances for the ontology concepts $C_1$ ... $C_n$)

Here $C \in \zeta$ | multiple assertions of (*someValuesFrom* ($\exists$) of $P_s$ *has* ($C_1, C_2 ... C_n$)) $\cap$ (*hasValue* ($\ni$) of $P_h$ *has* ($I_{c1}, I_{c2} ..... I_{cn}$)) imply:

If R = Database relation/view
and S =
$\{(P_S = C_1 \wedge P_h = I_{c1}), (P_S = C_2 \wedge P_h = I_{c2}), ....., (P_S = C_n \wedge P_h = I_{cn})\}$
then
$Q_R = R \div$ (*divided by*) S
$Q_R = \{t [a_1, a_2 ... a_n] :$
$t \in R \wedge \forall s \in S((t[a_1,....., a_n] \cup s) \in R)\}$

Example:

If R = DB relation/view
i.e. R = patients_information (patient_id, clinical_ test_ name, clinical_test_value)

And
$$S = \{(clinical\_test\_name = double\_vision \wedge$$
$$clinical\_test\_value = true),$$
$$(clinical\_test\_name = headaches \wedge$$
$$clinical\_test\_value = true),$$
$$(clinical\_test\_name = orthopedic\_sequelea \wedge$$
$$clinical\_test\_value = severe\_symptomatic)\}$$

So with respect to R the S is
$$\{(double\_vision, true), (headaches, true),$$
$$(orthopedic\_sequelea, severe\_symptomatic)\}$$

Then $Q_R = R \div (divided\ by)\ S$

As mentioned earlier in this paper, the ontology-to-database mapping information is stored within the ontology server, which includes the information about ontology property links, database name, table names, column names, primary and foreign keys. Once the query reformulation engine transforms the DL constructs into respective relational constructs, the ontology property information is updated with the database information. Finally, the reformulated relational query is passed to the query processing engine for execution.

Although the SQL relational algebra operations cover many cases as specified above, there are situations in which some additional translations are required. For example, matching for different date formats, partial string matching etc.; these are not covered in this paper but will be considered in future work. Regarding database *join* operations, until now we have considered only the *natural join* operation and have not dealt directly with the *theta, semi* and *outer join* operations between the database tables. For these join operations and relational algebra set operators, database views have been used to test the translations.

This approach has been applied on a part of the integrated HeC patients' database schema along with the implementation of an extensive graphical user interface (GUI) to perform query formulation and reformulation tasks. Due to scope and space limitations, detailed GUI descriptions have not been discussed in this paper. The prototype system has been presented to the HeC consortium and domain experts who have confirmed its potential functionality.

The current work in the project centres around evaluating the correctness of the above translation heuristics applied to a larger data-set and to extend the query reformulation algorithms, where necessary.

## 4. Conclusions

The central aim of this work was to provide the end users and their applications with query reformulation services using a domain ontology, with the main task of generating relational queries without requiring a complete knowledge of the information structure and access mechanisms of the underlying data sources. This involved the design of a query reformulation architecture with two main layers, the *ontological knowledge interface* and the *query reformulation engine* respectively.

The task of query reformulation has been automated by the successive incremental development of algorithms, to test the extent to which this procedure could be effectively automated. One of the key merits of this approach is that no interpretation of data needs to be carried out to be stored as ontology instances. This is clearly beneficial since the interpretation of data in existing data source(s) may cause some serious scalability issues with existing legacy applications. Secondly, it does not require its users to be familiar with the overall contents of the ontology to generate queries. This is helpful for the users who do not fully understand the system; navigating in large ontologies to select appropriate terms can itself be problematic. Moreover, the ontological information is accessed from the ontology server through customized wrapper methods, which is favorable while using large domain ontologies. Furthermore, the *query reformulation engine* is composed of generic Description Logic to Relational Query translation algorithms, and therefore can be easily employed for other domains.

While the implemented rules to translate OWL-DL queries to respective relational queries are heuristic based, further work is being carried out in the context of the HeC project to provide a formal ground to translate from description logic based ontologies to relational queries. The latter work will enable us to formally inform the verifiability of these anticipated translations from a point of view of correctness, consistency, and completeness. Also, there are issues that remain to be handled when using this heuristic approach. This is especially true when establishing the order and combinations of ontological expressions before they can be translated to relational queries. Despite these limitations, the current research work has provided us with a deeper insight into the problem by formulating a set of heuristics as a step to guiding the anticipated automation of this ontology-relational translation process. Finally, we anticipate that this approach will pave the way for a reflective process where results of queries' execution will enrich the current repository of domain ontologies.

## Acknowledgments

The authors would like to acknowledge the support and assistance of all partners in the Health-e-Child project, with special thanks to colleagues at CERN, Geneva and colleagues at UWE, Bristol in particular to Dr Peter Bloodsworth.